\documentclass[prd,twocolumn,floatfix,preprintnumbers,showpacs,letterpaper]{revtex4}
\usepackage{graphics}
\usepackage{epsfig}
\usepackage{amsmath}

\newcommand{\fracd}[1]{\frac{\dot{#1}}{#1}}

\def\GeV{{\rm GeV}}
\def\cm{{\rm cm}}
\def\vrel{v_{\rm rel}}

\begin{document}

\title{Cosmic microwave background and large scale structure limits on
the interaction between dark matter and baryons}
\author{Xuelei Chen}
\email{xuelei@itp.ucsb.edu}
\affiliation{Institute for Theoretical Physics, 
University of California, Santa Barbara, CA~~93106, USA}
\author{Steen Hannestad}
\email{steen@nordita.dk}
\affiliation{NORDITA, Blegdamsvej 17, DK-2100 Copenhagen, Denmark}
\author{Robert J. Scherrer}
\email{scherrer@pacific.mps.ohio-state.edu}
\affiliation{Department of Physics and Department of Astronomy, 
The Ohio State University,
Columbus, OH~~43210, USA}
\date{{February 26, 2002}}

\preprint{NSF-ITP-02-13}

\begin{abstract}

We study the effect on the cosmic microwave background (CMB)
anisotropy and large scale structure (LSS) power spectrum 
of a scattering interaction between cold dark matter and baryons.  
This scattering alters the CMB anisotropy and LSS spectrum
through momentum transfer between the cold dark matter particles 
and the baryons.
We find that current CMB observations can put an upper limit on the
scattering cross section which is comparable with 
or slightly stronger than previous disk heating
constraints at masses greater than 1 GeV, and much stronger at smaller masses.
When large-scale structure constraints are added to the CMB limits,
our constraint is more stringent than this previous limit at all masses.
In particular, a dark matter-baryon scattering cross section comparable to
the ``Spergel-Steinhardt'' cross section is ruled out for dark matter mass 
greater than 1 GeV.
\end{abstract}
\maketitle

\section{Introduction}
It is now generally accepted that the energy density of the universe
includes a substantial fraction of cold dark matter (CDM), constituting
roughly $0.3 - 0.4$ of the closure density.  It is normally assumed
that the CDM particle does not interact with ordinary baryonic matter
or with photons; hence the ``dark'' in cold dark matter.  However, a small
scattering cross section cannot be entirely ruled out, only constrained
by observations.

Several different models are possible for interactions between
ordinary matter and CDM.  One possibility is strongly interacting dark
matter (also known as strongly interacting massive particles, or SIMPs),
in which the CDM particle couples to baryons, but not to
photons or electrons. Recently, to solve the small scale problem in 
CDM models, strongly self-interacting particles were suggested
\cite{Spergel,Wandelt}, with
\begin{equation}
\label{eq:Spergel-Steinhardt}
8\times 10^{-25} < \frac{\sigma/\cm^2}{m/\GeV} < 10^{-23}.  
\end{equation}
The range quoted here (from 
Ref. \cite{Wandelt}) is narrower than originally proposed in
\cite{Spergel} because of additional constraints. While the original 
Spergel-Steinhardt cross section applies only to dark matter
self-interaction, if such an interaction is mediated by the strong force, then 
the dark matter might also interact strongly with
baryons\cite{Wandelt}. Examples of
such dark matter includes, e.g., quark-gluino bound
states \cite{Farrar}, strangelets \cite{Jaffe}, gauge singlet 
mesons \cite{Bento}, and Q-balls \cite{Kusenko}. 
Limits on such interactions were investigated
by Starkman et al. \cite{Starkman}, and recently reexamined 
in references \cite{Wandelt,QW,Moh,Cyburt}.
A second possibility is to
couple the dark matter particle electromagnetically, e.g., by giving it a tiny
electromagnetic charge (see, e.g., reference \cite{millicharge} and
references therein); such a CDM particle will scatter primarily off of
photons and free electrons.

Any interactions of this sort will affect the spectrum of
the cosmic microwave background (CMB) fluctuations, since they
act to transfer momentum from the dark matter to the baryon-photon
fluid, although for WIMPs this effect is negligible \cite{Chen,Hofmann}.
Current CMB observations can place constraints on such models.
The case of direct
interactions between the CDM particle and photons has already been
examined \cite{Boehm}, so we do not consider it here.  While
the model discussed in reference \cite{Boehm} is not identical
to the case of electromagnetically charged dark matter, since
it includes photon-CDM interactions but not electron-CDM interactions,
we expect it to be qualitatively similar to the charged dark matter model.
Therefore, we consider only the case of strongly-interacting dark matter,
in which the dark matter particle couples to baryons (only).

In the next section, we give the perturbation equations
for CDM which couples to baryons, and we show the matter
transfer function and CMB power spectra for some representative cases.
In Sec. 3, we compare our calculations with the CMB observations to derive
constraints on the cross section for dark matter-baryon scattering,
and we compare to previously-derived limits.  Our conclusions are summarized
in Sect. 4.

\section{Fluctuation growth with strongly-interacting dark matter}

\subsection{The Basic Modified Equations}
We begin by assuming a scattering cross section $\sigma$ between
the CDM particle of mass $m_c$ and protons with mass $m_b$,
and we derive the changes to the perturbation equations produced by
this scattering interaction.
Our starting point is the set of perturbation equations in
reference \cite{MB95}, which form the basis for CMBFAST
\cite{CMBFAST}.  We will not reproduce here all of the perturbation
equations, but simply give those equations which are modified.

Following reference \cite{MB95}, we work in the synchronous
gauge.  However,  because the CDM is interacting it does not provide 
a natural way of defining the synchronous coordinates, as opposed
to standard CDM \cite{MB95}.
Let $\delta_c$ and $\delta_b$ be the
density fluctuation ($\equiv \delta \rho/\rho$) for the cold dark
matter and the baryons, respectively, and let $\theta_c$ and
$\theta_b$ be the corresponding velocity divergences.
Then for the cold dark matter, we have
\begin{eqnarray}
\dot{\delta}_c &=& -\theta_c - \frac{1}{2} \dot{h}, \\
\dot{\theta}_c &=& -\fracd{a} \theta_c + c_c^2 k^2 \delta_b
+ R_c
\end{eqnarray}
For baryons the corresponding terms are
\begin{eqnarray}
\dot{\delta}_b &=& -\theta_b - \frac{1}{2} \dot{h}, \\
\dot{\theta}_b &=& 
-\fracd{a} \theta_b + c_b^2 k^2 \delta_b + R_{\gamma}+ R_b.
\end{eqnarray}
with 
\begin{eqnarray}
R_b &\equiv&  K_c (\theta_b - \theta_c) = 
\frac{m_c n_c \sigma \vrel a}{m_b+m_c} (\theta_b - \theta_c),\\ 
\nonumber\\
R_c &\equiv& K_b  (\theta_c - \theta_b) = 
\frac{m_b n_b \sigma \vrel a}{m_b+m_c}(\theta_c - \theta_b),\\
\nonumber\\
R_{\gamma} &\equiv& K_{\gamma}(\theta_{\gamma}-\theta_b)
=\frac{4\rho_{\gamma}}{3\rho_b} a n_e \sigma_T(\theta_{\gamma}-\theta_b)
\end{eqnarray}
In these equations $c_c$ and $c_b$ are the adiabiatic sound speed
of the CDM and baryons, respectively, $\vrel$ is
the average relative velocity between the CDM and the baryons, and
$R_c, R_b, R_{\gamma}$ are the momentum transfer terms.  
The sound speeds are given by
\begin{eqnarray}
c_b^2 &=& (k_B T_b/\mu) 
\left(1-\frac{1}{3} \frac{d\ln T_b}{d\ln a}\right), \\
c_c^2 &=& (k_B T_c/m_c) 
\left(1-\frac{1}{3} \frac{d\ln T_c}{d\ln a}\right),
\end{eqnarray}
the average relative velocity by
\begin{equation}
\vrel = \sqrt{\frac{3(m_c T_b + m_b T_c)}{m_b m_c}},
\end{equation}
In calculating average velocities we have assumed thermal Maxwell-Boltzmann
distributions for both CDM and baryons. The temperature evolution
of the dark
matter and baryons is given by
\begin{eqnarray}
\dot{T}_b + 2\fracd{a} T_b &=& 
-K_{\gamma} (T_b-T_{\gamma})-K_c (T_b-T_c),\\
\dot{T}_c + 2\fracd{a} T_c &=& -K_b (T_c-T_b).
\end{eqnarray}

It is possible to obtain analytical solutions for the temperature 
evolution equations, but they are
complicated and cumbersome to use. Instead, we use 
a semi-implicit scheme to integrate these differential equations.
We tested this method in a few cases, 
and found that it produces solutions which agree 
with the analytical solutions
within 0.2\%, and the effect on the CMB spectrum is negligible.

\subsection{The Effect of Primordial Helium}

This set of equations ignores the fact that the baryonic fluid
consists of two species:  hydrogen and helium-4, in a roughly 3 to 1
ratio by mass.  To account for the presence of helium as well
as hydrogen, we make the following replacements:
\begin{widetext}
\begin{eqnarray}
\frac{m_b n_b \sigma\vrel}{m_b+m_c}  &\rightarrow& 
\frac{m_{\rm H} n_{\rm H} \sigma_{\rm H} v_{\rm H}}{m_{\rm H}+m_c} 
+\frac{m_{\rm He} n_{\rm He} \sigma_{\rm He} v_{\rm He}}{m_{\rm
He}+m_c}, \\
\frac{m_c n_c \sigma\vrel }{m_b+m_c}  &\rightarrow& 
 \frac{m_c n_c \sigma_{\rm H}v_{\rm H}}{m_{\rm H}+m_c}(1-Y)  + 
\frac{m_c n_c \sigma_{\rm He}v_{\rm He}}{m_{\rm He}+m_c}Y,
\end{eqnarray}
\end{widetext}
where $Y$ is the primordial mass fraction of helium-4 (we
take $Y = 0.24$ throughout), and we define
separate velocities for the hydrogen and helium relative to the dark matter:

\begin{eqnarray}
v_{\rm H}&=&\sqrt{\frac{3(m_c T_b + m_{\rm H} T_c)}{m_{\rm H} m_c}}\\
v_{\rm He}&=&\sqrt{\frac{3(m_c T_b + m_{\rm He} T_c)}{m_{\rm He} m_c}}.
\end{eqnarray}

In our calculations, $\sigma$, the scattering cross section
between the CDM particle and the proton, is
a free parameter which we will constrain from the observations.
However, we need to make some sort of assumption regarding
$\sigma_{He}$, the scattering cross section between CDM and helium
nuclei.  For coherent scattering, we expect $\sigma_{He} = 16 \sigma$,
while a spin-dependent cross section would yield $\sigma_{He} = 0$.
For now, we will examine the
intermediate case of incoherent scattering,
for which $\sigma_{He} = 4 \sigma$.  Later,
in Sec. 3, we will extend our calculation to consider both coherent
scattering and spin-dependent scattering.

Taking $m_{\rm He} \simeq 4m_{\rm H}$, and
$n_b = \rho_b/ m_{\rm H}$, 
the above substitution can be achieved by simply replacing everywhere
\begin{widetext}
\begin{eqnarray}
\frac{m_b \sigma \vrel}{m_b+m_c}  &\longrightarrow&
\frac{m_{\rm H} \sigma}{m_{\rm H}+m_c}
\left[(1-Y) v_{\rm H}+ Y\left(\frac{\sigma_{\rm He}}{\sigma_{\rm H}}\right) 
\frac{m_{\rm H}+ m_c}{4m_{\rm H}+m_c} v_{\rm He}\right]\\
\frac{m_c \sigma \vrel}{m_b+m_c}  &\longrightarrow&
\frac{m_c \sigma}{m_{\rm H}+m_c}
\left[(1-Y) v_{\rm H}+ Y \left(\frac{\sigma_{\rm He}}{\sigma_{\rm H}}\right)
\frac{m_{\rm H}+ m_c}{4m_{\rm H}+m_c} v_{\rm He}\right].
\end{eqnarray}
\end{widetext}

\subsection{Tight-Coupling Approximation}

Following reference \cite{MB95},
we use a separate set of equations in the limit of tight coupling
between baryons and photons. 
In the tight-coupling regime, the temperature evolution is given by
\begin{widetext}
\begin{eqnarray}
\dot{T}_b &=& -2\fracd{a} T_b + \frac{8}{3}\frac{\mu}{m_e} 
\frac{\rho_{\gamma}}{\rho_b} a n_e \sigma_T (T_{\gamma}-T_b) +
K_c (T_c -T_b)\\
\dot{T}_c &=& -2\fracd{a} T_c  +
K_b (T_b -T_c)
\end{eqnarray}
\end{widetext}
where $\mu$ is the mean molecular weight.
During tight coupling, $T_c \sim T_b \sim T_{\gamma}$,
$\dot{T_b} \sim -\frac{\dot{a}}{a} T_b$, and
$\dot {v}_{\rm rel} = -\frac{1}{2} \fracd{a} \vrel$.
Setting $R=4\rho_{\gamma}/3\rho_b$, and 
$\tau_{\gamma} = (a n_e \sigma_T)^{-1}$, we have
\begin{widetext}
\begin{equation}
\dot{R}_b+\fracd{a}R_b=\frac{m_c n_c \sigma a \vrel}{m_b + m_c}
\left[(\dot\theta_c-\dot\theta_b)-\frac{3}{2}\fracd{a}(\theta_c-\theta_b)\right]
\end{equation}
\begin{equation}
(1+R)\dot{\theta_b} +\fracd{a} \theta_b - c_b^2 k^2 \delta_b 
-k^2 R\left(\frac{1}{4}\delta_{\gamma}-\sigma_{\gamma}\right) 
+ R(\dot\theta_{\gamma}-\dot\theta_b)-R_b=0,
\end{equation}
\begin{equation}
\theta_b -\theta_{\gamma} = \frac{\tau_{\gamma}}{1+R}\left[-\fracd{a}\theta_b
+k^2 (c_b^2 \delta_b -\frac{1}{4}\delta_{\gamma}+\sigma_{\gamma})
+\dot\theta_{\gamma}-\dot\theta_b+R_b \right],
\end{equation}
\begin{equation}
\dot\theta_b-\dot\theta_{\gamma} =
\frac{\tau_{\gamma}}{1+R}
\left[-\frac{\ddot{a}}{a}\theta_b -\fracd{a} k^2
\frac{\delta_{\gamma}}{2} + k^2 \left(c_b^2 \dot\delta_b -\frac{1}{4}
\dot\delta_{\gamma}\right)+\fracd{a}R_b + \dot{R}_b \right] +
\frac{2R}{1+R}\fracd{a}\left(\theta_b -\theta_{\gamma}\right).
\end{equation}
\end{widetext}

\subsection{Applicable range of our calculation}
In the calculation described above, we have assumed that the
dark matter is made of a single type of particle, with density
parameter $\Omega_m$.
We have also assumed that it is
a ``cold dark matter'' particle, i.e. during the whole range of our
calculation it is non-relativistic, which translates to 
$m > 1$ MeV. We have also assumed that during each scattering 
only one baryon interacts with the dark matter particle. 
If the interaction length of the particle is 
$a$ ($\sigma \equiv 4\pi a^2$), then we require that this interaction 
length be less than the inter-particle spacing:
\begin{equation}
a < n_b^{-1/3},
\end{equation}
where $n_b \sim 10^{-5} (1+z)^3 h_0^2 \Omega_b \cm^{-3}$. 
Processes which affect CMB anisotropy happen at redshift less than 
$10^5$, so we can put a limit on the applicable cross section, neglecting 
factors of order unity:
\begin{equation}
\sigma_{\rm max} \sim 10^{-6} \cm^2 .
\end{equation}
Our calculation must be modified for $\sigma>\sigma_{\rm max} $ to
take into account multi-particle scattering.

\subsection{The matter and CMB power spectra}

We have modified the CMBFAST code \cite{CMBFAST} to include the effects
discussed in the previous sections.  For initial conditions, we
assume that the dark matter and baryons started tightly coupled
with the same temperature.  Tensor modes have been ignored, and
we consider only flat geometries ($\Omega = 1$).

\begin{figure}[htbp]
\begin{center}
\epsfig{file=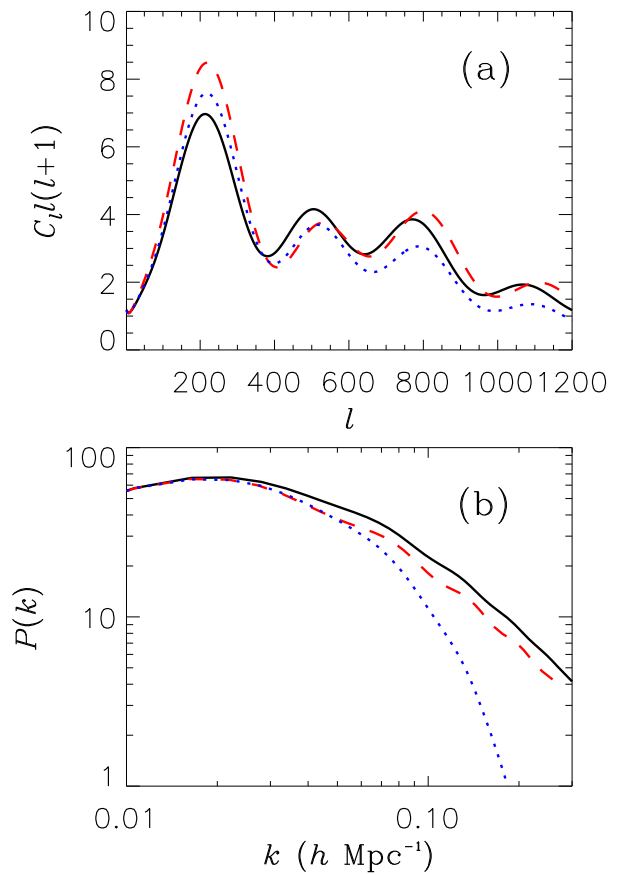,width=0.5\textwidth}
\end{center}
\vspace*{1.0cm}

\caption{\label{fig:cl} (a) CMB power spectra for a few sample cases.
The solid (black) curve is a fiducial $\Lambda$CDM model with 
parameters $\Omega=1$, $\Omega_b h^2 = 0.018$, $\Omega_m h^2 = 0.18$,
$H_0 = 75 \,\, {\rm km} \, {\rm s}^{-1} \, {\rm Mpc}^{-1}$, $n=1$,
and without dark matter-baryon interaction. The dotted (blue) curve is
for a dark matter mass of 1 GeV, with a dark matter-baryon cross
section $\sigma = 3 \times 10^{-24} \,\, {\rm cm}^2$.  The dashed
(red) curve is for a model without dark matter-baryon interaction, but
with increased baryon density, $\Omega_b = 0.05$.  All spectra are
calculated without COBE-normalization (arbitrary units) and
taking $\sigma_{He} = 4\sigma$. (b) Matter
power spectra for the same models as in (a). Normalization is
arbitrary.}
\end{figure}

In Fig.~\ref{fig:cl} we show the CMB fluctuation spectra
and the corresponding square of the matter transfer function
for some representative cases.  The coupling of the dark matter
to the baryons damps the matter power spectrum on small scales,
as can be seen in Fig. ~\ref{fig:cl}(b).

One might imagine that the effect on the CMB of 
coupling the baryons to the dark matter
would be equivalent to a standard model with a larger value of $\Omega_b$.
This is not the case, as can be seen in Fig.~\ref{fig:cl}(a).  
Although both
models produce an increase in the amplitude of the first acoustic
peak, the large $\Omega_b$ model produces an increase in the amplitude
of the third
peak, while the strongly-interacting dark matter model
yields a decrease in the third peak amplitude (and in all
of the other peaks as well).

One can understand the difference between the
CMB spectra in the following way: The 
baryon-photon oscillations are damped oscillations, with a damping
rate (viscosity) inversely proportional to the 
baryon-photon interaction rate. Interacting CDM with intermediate 
cross section amounts to adding ``baryons'' but simultaneously also increasing
the viscosity of the plasma. Therefore the acoustic oscillations damp
quickly and all peaks but the first are suppressed. 
Scales around the first acoustic peak have only undergone part of an 
oscillation and therefore have not been damped significantly. On such
large scales the increase in oscillation amplitude due to added
``baryons'' is more important and the overall power spectrum increases.

On the other hand, increasing the baryon density has the well-known effect of
increasing the height of the compression peaks (1st, 3rd,...) and
decreasing the height of the decompression peaks (2nd, 4th,...).
Of course, for very large cross sections, the effective viscosity of the
CDM-baryon-photon plasma approaches that of a pure baryon-photon plasma
and the CDM acts exactly like baryons.
An effect similar to the increased viscosity due to CDM-baryon
interactions can be seen in models with increased width
of the last scattering surface due to non-standard recombination,
where the effective viscosity of the baryon-photon plasma also increases 
around the time of recombination \cite{HS01}.

The effect on the matter power spectrum seen in Fig.~\ref{fig:cl}(b)
can be understood in the same way. The increased viscosity in
the interacting model leads to a much stronger damping on small scales
than in the model with increased baryon density.

\section{Comparison with observations}

In order to constrain models with CDM-baryon interactions we
compare the CMB and matter power spectra of such models with 
recent observational data.

{\it CMB data set ---}
Several data sets of high precision are now publicly available.
In addition to the COBE \cite{Bennett:1996ce} data for small 
$l$ there are data from
BOOMERANG \cite{boom}, MAXIMA \cite{max}, DASI \cite{dasi} 
and several other experiments \cite{WTZ,qmask}.
Wang, Tegmark and Zaldarriaga \cite{WTZ}
have compiled a combined data set
from all these available data, including calibration errors.
In the present work we use this compiled data set, which is both
easy to use and includes all relevant present information.

{\it LSS data set ---}
At present, by far the largest survey available is the 2dF
\cite{2dF} of which about 147,000 galaxies have so far been
analysed. Tegmark, Hamilton and Xu \cite{THX} have calculated a power
spectrum, $P(k)$, from this data, which we use in the present work.
The 2dF data extends to very small scales where there are large
effects of non-linearity. Since we only calculate linear power spectra,
we use (in accordance with standard procedure) only data on scales larger than
$k = 0.2 h \,\, {\rm Mpc}^{-1}$, where effects of non-linearity
should be minimal.

The CMB fluctuations are usually
described in terms of the power spectrum, which is again expressed in
terms of $C_l$ coefficients as $l(l+1)C_l$, where
\begin{equation}
C_l \equiv \langle |a_{lm}|^2\rangle.
\end{equation}
The $a_{lm}$ coefficients are given in terms of the actual temperature
fluctuations as
\begin{equation}
T(\theta,\phi) = \sum_{lm} a_{lm} Y_{lm} (\theta,\phi).
\end{equation}
Given a set of experimental measurements, the likelihood function is
\begin{equation}
{\cal L}(\Theta) \propto \exp \left( -\frac{1}{2} x^\dagger
[C(\Theta)^{-1}] x \right),
\end{equation}
where $\Theta = (\Omega, \Omega_b, H_0, n, \tau, \ldots)$ is a vector
describing the given point in parameter space, $x$ is a vector
containing all the data points, and $C(\Theta)$ is the data covariance
matrix.  This applies when the errors are Gaussian. If we also assume
that the errors are uncorrelated, it can be reduced to the simple
expression, ${\cal L} \propto e^{-\chi^2/2}$, where
\begin{equation}
\chi^2 = \sum_{i=1}^{N_{\rm max}} \frac{(C_{l, {\rm obs}}-C_{l,{\rm
theory}})_i^2} {\sigma(C_l)_i^2},
\label{eq:chi2}
\end{equation} 
is a $\chi^2$-statistic and $N_{\rm max}$ is the number of power
spectrum data points \cite{oh}.  In the present letter we use
equation (\ref{eq:chi2}) for calculating $\chi^2$.
In the case where we also use LSS data $\chi^2$ is instead given by
\begin{widetext}
\begin{equation}
\chi^2 =  \sum_{i=1}^{N_{\rm max,CMB}} \frac{(C_{l, {\rm obs}}-C_{l,{\rm
theory}})_i^2} {\sigma(C_l)_i^2} ~~~+ 
\sum_{j=1}^{N_{\rm max,LSS}} \frac{(P(k)_{{\rm obs}}-P(k)_{{\rm
theory}})_j^2} {\sigma(P(k))_j^2}.
\label{eq:chi22}
\end{equation} 
\end{widetext}

The procedure is then to calculate the likelihood function over the
space of cosmological parameters. 
The 2D likelihood function for $(m_c,\sigma)$ is obtained by keeping
$(m_c,\sigma)$ fixed and marginalising over all other parameters.

As free parameters in the likelihood analysis we use
$\Omega_m$, the matter density, $\Omega_b$, the baryon density,
$H_0$, the Hubble parameter, $n$, the scalar spectral index,
$\tau$, the optical depth to reionization, and $Q$, the overall
normalization of the data. When large scale structure constraints
are included we also use $b$, the normalization of the matter
power spectrum, as a free parameter. This means that we treat
$Q$ and $b$ as free and uncorrelated parameters. This is very
conservative and eliminates any possible systematics involved in
determining the bias parameter.
We constrain the analysis to flat ($\Omega_m + \Omega_\Lambda = 1$)
models, and we assume that the tensor mode contribution is 
negligible ($T/S=0$).
These assumptions are compatible with analyses of the present data
\cite{WTZ}, and relaxing them does not have a big effect on 
the final results.
For maximizing the likelihood function we use a simulated annealing
method, as described in reference~\cite{Hannestad:2000wx}.

Table I shows the different priors used. In the ``CMB'' prior 
the only important constraint is that $0.4 \leq h \leq 0.9$
($h \equiv H_0/(100 \,\, {\rm km} \, {\rm s}^{-1} \, {\rm Mpc}^{-1})$).
For the CMB+$H_0$+BBN prior we use the constraint 
$H_0 = 72 \pm 8 \,\, {\rm km} \, {\rm s}^{-1} \, {\rm Mpc}^{-1}$
from the HST Hubble key project \cite{freedman} (the constraint is added
assuming a Gaussian distribution) and the constraint $\Omega_b h^2
= 0.020 \pm 0.002$ from BBN \cite{Burles:2000zk}.
Finally, in the $H_0$+BBN+LSS case, we add data from the 2dF survey
\cite{THX}.

\begin{figure}[htbp]
\begin{center}
\epsfig{file=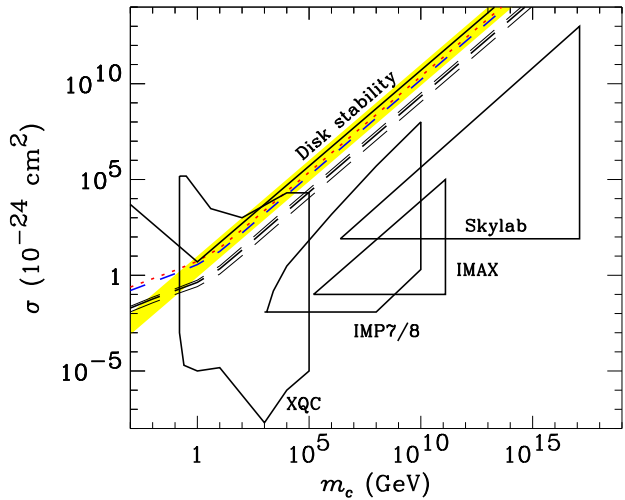,width=0.5\textwidth}
\end{center}
\bigskip
\caption{\label{fig:excl} 
Excluded regions in the $(m_c,\sigma)$ plane for
strongly-interacting dark matter, where $m_c$
is the mass of the dark matter particle, and $\sigma$
is the cross section for scattering between dark matter
and protons.  Dashed and dotted lines
are the CMB limits derived in this
paper with various priors: dotted (red) is CMB only, short-dash (blue)
is CMB+BBN+$H_0$, thick long-dash (black)
is CMB+BBN+$H_0$+LSS.  All of these limits
assume a scattering cross-section between helium
and dark matter of $\sigma_{He} = 4 \sigma$.
Thin dashed lines give the limits for CMB+BBN+$H_0$+LSS
for spin-dependent scattering between helium and dark matter
($\sigma_{He} = 0$, upper curve) and coherent scattering
($\sigma_{He} = 16 \sigma$, lower curve).
In all of these cases,
the region above the curve is excluded at the 95\% confidence level.
Solid lines give excluded regions from other papers.
The region above the top solid line is excluded by the disk heating
constraint from Starkman {\it et al.} 
Closed solid contours give excluded regions from the indicated
space and balloon-based experiments discussed by Wandelt {\it et al.}
The regions enclosed by these curves are excluded.  The shaded (yellow) strip
is the region suggested in Wandelt {\it et al.} for which self-interacting
dark matter would have a comparable cross section for scattering off of
both baryons and itself.}
\end{figure}

Following Ref.~\cite{Wandelt}, we plot the constraints from various
experiments in the $(m_c,\sigma)$ plane in Fig.~\ref{fig:excl}; limits
obtained with different priors are plotted with different line styles.
The shaded region is the ``Spergel-Steinhardt'' region as given in
equation~(\ref{eq:Spergel-Steinhardt}); note that it is actually narrower than 
plotted in Fig.~3 of Ref.~\cite{Wandelt}.
This figure clearly shows that constraints from the CMB data alone
for large CDM mass ($m_c > m_b \sim 1$ GeV)
are comparable to limits from galactic disk
heating arguments \cite{Starkman}. 
On the other hand, for small CDM mass ($m_c < m_b \sim 1$ GeV),
the limit from the CMB alone is much stronger.  This difference
arises because the limits from disk heating are related to
energy transfer, which  
decreases as $(m_c/m_b)^2$ for $m_c  \ll m_b$, while the limits
from the CMB and large scale structure are based on
momentum transfer, which decreases only as $m_c/m_b$ for $m_c \ll m_b$.
At the same time, the relative velocity also increases for low 
CDM mass (as $m_c^{-1/2}$). This is the reason that, for $m_c \ll m_b$,
our bound goes as $m_c^{1/2}$, but for the disk heating argument
it goes as $m_c^{-1}$.
Our bound is therefore much stronger at low mass, even if only CMB
data is used.

The addition of priors on $\Omega_b$ and $H_0$ does not increase the
exclusion region by much, because there is very little degeneracy
between $(m_c,\sigma)$ and these two parameters.
Only when LSS data is added do we find a significant improvement.
The reason is that CDM-baryon interactions lead to a significant
suppression of small scale power, as discussed in the previous section.
Even for relatively small cross sections this effect leads to a 
disagreement with the 2dF data.  With the inclusion of LSS data, our
bound becomes much stronger than the
disk heating argument of Starkman {\it et al.} \cite{Starkman}, even
at large masses.  A good analytical fit to our 95\% C.L. limit is
\begin{equation}
\label{fit}
\frac{\sigma}{10^{-24} {\rm cm^2}} <\frac{0.63 x^{1/2} + 0.22 x^{3/2}}{1+x^{1/2}}
\end{equation}
where $x = m_c$/GeV.  Equation (\ref{fit}) is accurate to within 10\%.
For this case, we also consider the effect of coherent ($\sigma_{He} = 16\sigma$)
and spin-dependent ($\sigma_{He} = 0$) scattering between the dark matter and
helium.  The limits for these two cases are shown in Fig. 2 as thinner
dashed curves above (for $\sigma_{He} = 0$) and below (for $\sigma_{He} =
16\sigma$) the limit for $\sigma_{He} = 4 \sigma$.  Altering $\sigma_{He}$
in this way changes the upper bounds by a factor of roughly $2-3$, which
is barely noticeable on the scale of this graph.

\section{Conclusions}

CMB observations constrain any scattering interaction between
dark matter and baryons.  Our results indicate that the limit
from current CMB observations is comparable to previous limits
from disk heating \cite{Starkman} for masses greater than 1 GeV, and
the CMB limit is much stronger at smaller masses.  If we also
include large-scale structure data, then our limit
is more stringent that the disk-heating limit at all masses.
An analytical fit of our combined CMB+LSS limit is given in Eq.~(\ref{fit}).

Our CBM+LSS limits exclude the region discussed in Ref. \cite{Wandelt},
in which self-interacting dark matter interacts with baryons
with roughly the same cross section with which it interacts
with itself (Fig. 2).  This, by itself, does not exclude the
self-interacting dark matter scenario, since the dark matter
is not required to interact with baryons at all, but it does exclude
models with the indicated dark matter-baryon scattering cross section.

While space and balloon-based experiments can provide tighter constraints
for specific regions in the parameter space of CDM mass and cross section,
(Fig. 2 and
reference \cite{Wandelt}), the combination of CMB and large-scale structure
appears to provide the best general upper limit on
the CDM-baryon scattering cross section for arbitrary CDM masses
(and it is much stronger than all other limits at low masses).
Our constraints are somewhat stronger for
coherent scattering from helium nuclei and weaker
for a spin-dependent interaction.
Perhaps more importantly, these limits will only get better with
new CMB data.

\acknowledgments
X.C. is supported by the NSF under grant PHY99-07949, 
R.J.S. is supported by the DOE under grant DE-FG02-91ER40690.

\newcommand\AJ[3]{~Astron. J.{\bf ~#1}, #2~(#3)}
\newcommand\APJ[3]{~Astrophys. J.{\bf ~#1}, #2~ (#3)}
\newcommand\apjl[3]{~Astrophys. J. Lett. {\bf ~#1}, L#2~(#3)}
\newcommand\ass[3]{~Astrophys. Space Sci.{\bf ~#1}, #2~(#3)}
\newcommand\cqg[3]{~Class. Quant. Grav.{\bf ~#1}, #2~(#3)}
\newcommand\mnras[3]{~Mon. Not. R. Astron. Soc.{\bf ~#1}, #2~(#3)}
\newcommand\mpla[3]{~Mod. Phys. Lett. A{\bf ~#1}, #2~(#3)}
\newcommand\npb[3]{~Nucl. Phys. B{\bf ~#1}, #2~(#3)}
\newcommand\plb[3]{~Phys. Lett. B{\bf ~#1}, #2~(#3)}
\newcommand\pr[3]{~Phys. Rev.{\bf ~#1}, #2~(#3)}
\newcommand\PRL[3]{~Phys. Rev. Lett.{\bf ~#1}, #2~(#3)}
\newcommand\PRD[3]{~Phys. Rev. D{\bf ~#1}, #2~(#3)}
\newcommand\prog[3]{~Prog. Theor. Phys.{\bf ~#1}, #2~(#3)}
\newcommand\RMP[3]{~Rev. Mod. Phys.{\bf ~#1}, #2~(#3)}

\begin{center}
\begin{table*}[hb]
\caption{The different priors used in the analysis. Notice that in the
first two cases, the bias parameter $b$ is not used because
LSS data is not included in the fit.\\}

\begin{tabular}{lccccccc}
prior type & $\Omega_m$ & $\Omega_b h^2$ & $h$ & $n$ & $\tau$ & $Q$ & $b$ \cr
\colrule
CMB & $\Omega_b$-1 & 0.008 - 0.040 & 0.4-1.0 & 0.66-1.34 & 0-1 & free & not used \cr
CMB + BBN + $H_0$ & $\Omega_b$-1 & $0.020 \pm 0.002$ & $0.72 \pm 0.08$ & 0.66-1.34 
& 0-1 & free & not used  \cr
CMB + BBN + $H_0$ + LSS & $\Omega_b$-1 & $0.020 \pm 0.002$ & $0.72 \pm 0.08$ & 0.66-1.34 
& 0-1 & free & free \cr
\end{tabular}
\end{table*}
\end{center}

\end{document}